\documentclass[reprint,superscriptaddress, aps, pra, floatfix]{revtex4-1}

\usepackage{blindtext}
    \usepackage{amsmath}
    \usepackage{makeidx}
    \usepackage{amsfonts}
    \usepackage{float}
    \usepackage[usenames,dvipsnames]{pstricks}
    \usepackage{subfigure}
    \usepackage{epsfig}
    \usepackage{pst-grad} 
    \usepackage{pst-plot} 
    \usepackage[colorlinks,hyperindex]{hyperref}
    \hypersetup
    {
        colorlinks,%
        citecolor=black,%
        linkcolor=black,%
        urlcolor=black,%
    }

\usepackage{color}
\usepackage{dsfont}
\usepackage{subfigure}

\usepackage{dcolumn}
\usepackage{amsmath}    
\usepackage{verbatim}   
\usepackage{color}      
\usepackage{hyperref}   
\usepackage{amsfonts}
\usepackage{braket}
\usepackage{bm}
\usepackage{epstopdf}
\usepackage{titlesec}
\usepackage{capt-of}
\usepackage{float}

\usepackage[a4paper, left=17mm, right=17mm, top=20mm]{geometry}

\begin{document}


\title{Towards a source of multi-photon entangled states for linear optical quantum computing
}

\author{J. P. Lee}\,
\thanks{These authors contributed equally}
\affiliation{Toshiba Research Europe Limited, Cambridge Research Laboratory, 208 Science Park, Milton Road, Cambridge, CB4 0GZ, U.K.}
\affiliation{Engineering Department, University of Cambridge, 9 J. J. Thomson Avenue, Cambridge, CB3 0FA, U.K.}
\email{james.patrick.lee.47@gmail.com}

\author{B. Villa}\,
\thanks{These authors contributed equally}
\affiliation{Toshiba Research Europe Limited, Cambridge Research Laboratory, 208 Science Park, Milton Road, Cambridge, CB4 0GZ, U.K.}
\affiliation{Cavendish Laboratory, Cambridge University, J. J. Thomson Avenue, Cambridge, CB3 0HE, U.K.}

\author{A. J. Bennett}
\altaffiliation{Present address: Institute for Compound Semiconductors, Cardiff University, Queen's Buildings, 5 The Parade, Roath, Cardiff, CF24 3AA, U.K.}
\affiliation{Toshiba Research Europe Limited, Cambridge Research Laboratory, 208 Science Park, Milton Road, Cambridge, CB4 0GZ, U.K.}

\author{R. M. Stevenson}

\author{D. J. P. Ellis}
\affiliation{Toshiba Research Europe Limited, Cambridge Research Laboratory, 208 Science Park, Milton Road, Cambridge, CB4 0GZ, U.K.}

\author{I. Farrer}
\altaffiliation{Present address: Department of Electronic \& Electrical Engineering, University of Sheffield, Mappin Street, Sheffield S1 3JD, United Kingdom}
\affiliation{Cavendish Laboratory, Cambridge University,\\
J. J. Thomson Avenue, Cambridge, CB3 0HE, U.K.}

\author{D. A. Ritchie}
\affiliation{Cavendish Laboratory, Cambridge University,\\
J. J. Thomson Avenue, Cambridge, CB3 0HE, U.K.}

\author{A. J. Shields}
\affiliation{Toshiba Research Europe Limited, Cambridge Research Laboratory, 208 Science Park, Milton Road, Cambridge, CB4 0GZ, U.K.}

\date{\today}%

\begin{abstract}

    We propose a scheme to make use of recent advances in cavity QED-enhanced resonance fluorescence from quantum dots to generate a stream of entangled and indistinguishable photons. We then demonstrate that we can optically manipulate the state of a trapped hole spin to achieve complete coherent control of a qubit. In combination with the selective cavity enhancement of the resonantly excited transition, we use this capability to perform a proof-of-principle demonstration of our proposal by showing that the time bin of a single photon is dependent on the measured state of the trapped spin. 
\end{abstract}

\maketitle

\section{Introduction}
    One of the primary difficulties with optical approaches to quantum computing comes from the weakness of the photon-photon interaction. Although this weak interaction means that photons suffer little from decoherence, it makes performing conditional operations on two photons a challenging endeavour \cite{hacker2016photon, PhysRevA.91.030301}. One possible approach, linear optical quantum computing, was shown to be a scalable by Knill, Laflamme and Milburn in 2000 \cite{knill2001scheme}. The scheme sidesteps the obstacle of interacting two-photons by making use of arrays of phase shifters and beam splitters, and measurement to induce effective photon-photon interactions probabilistically via the Hong-Ou-Mandel (HOM) effect. The scheme requires an almost infeasible number of optical components \cite{van2011optical}, in addition to reliable and indistinguishable single photon sources and efficient detectors. There have been proposals that reduce the experimental requirements by generating entangled states to be used as a resource for measurement based quantum computing \cite{nielsen2006cluster}, but these still remain impractical with current technology. 
    However, in light of the continuing advances in integrated photonic circuits \cite{murray2015quantum, wang2017experimental, zadeh2018single, marsili2013detecting, ellis2018independent}, it has been suggested that a source of three photon entangled states, suitable for stitching together by a HOM based fusion mechanism \cite{browne2005resource}, would finally bring the photonic quantum computer to within reach \cite{rudolph2017optimistic}.

    There have been multiple proposals for generating such entangled states \cite{PhysRevA.82.032332, economou2010optically, clark2010cluster}, but of particular interest here is Lindner and Rudolph's 2009 proposal for generating a chain of entangled photons from a singly charged quantum dot \cite{lindner2009proposal}. Their scheme involves the repeated resonant excitation of a quantum dot containing a single trapped charge obeying Faraday geometry-like selection rules.

\section{The experimental implementation of Lindner and Rudolph's scheme}
    The scheme relies on resonant excitation to generate the photons. Consequently, the resonant laser light that is reflected from the sample must be separated from the output signal as it cannot be spectrally filtered. This is normally done using polarisation filtering; the commercial availability of high quality polarisers means that signal-to-background ratios in excess of $10^{3}:1$ are routinely observed \cite{bennett2016cavity, kuhlmann2013dark}. However, as the entanglement is encoded in the polarisation of the photons, polarisation filtering would destroy it. Although there are other approaches to performing resonance fluorescence experiments without the use of polarisation optics \cite{muller2007resonance}, these approaches are still under development and require sophisticated fabrication techniques.

   A further key consideration involves the choice of magnetic field configuration. A Faraday magnetic field, where only the vertical transitions are allowed (cf. Fig. \ref{scheme}), is required for the sequential generation of a cluster state following the proposal in \cite{lindner2009proposal}. However, the authors suggest adding a weak Voigt magnetic field to allow the spin to precess in order to perform some of the operations needed to generate the cluster state. It is unclear to what extend this Voigt field would remove the Faraday geometry-like selection rules, as in a Voigt field both vertical and diagonal transitions are allowed. This is complicated by the fact that, ideally, one would use a large magnetic field, as this has been shown to improve the coherence time of the trapped spin \cite{stockill2016quantum}. In addition, Voigt magnetic fields have been exploited to demonstrate high fidelity coherent optical spin rotations, which has not been reproduced in Faraday fields. It would therefore be advantageous to modify the scheme to remove the necessity of a Faraday field altogether. Of particular interest here are the experiments reported in \cite{greilich2009ultrafast}, where the authors used a several picosecond circularly polarised pulse to induce a spin rotation via the AC Stark effect. This technique would allow the implementation of the spin rotations required to implement our proposal in a Voigt field.

    Schwartz {\it et al.} tackled these problems by using a dark exciton as the spin in the place of a trapped charge and resonantly driving the dark exciton-biexciton transition \cite{Schwartzaah4758}. The biexciton-dark exciton decay involves a radiative decay followed by a non-radiative decay, meaning that the emitted light differs in energy from the excitation light, allowing spectral filtering. It would be interesting to investigate how this process affects the coherence of the emitted light or whether it reduces the state fidelity by leaking information to the environment via phonons. It is worth noting that for a linear cluster state encoded in the photon polarisation, the coherence time of the emitted light does not matter. However, in order to be useful for quantum computing applications, the generated photons must be suitable for Hong-Ou-Mandel interference-based fusion operations, which necessitates good coherence properties because `the degree of indistinguishability equals the degree of coherence' \cite{Mandel:91}. 

    The work of Schwartz {\it et al.} proves the validity of Lindner and Rudolph's scheme and serves as motivation to explore the impact of applying recent developments in coherent spin control and cavity QED techniques to the challenge of building a source of photonic states useful for quantum computing applications.

\section{Proposal for generating time-bin-encoded entangled states}

    \begin{figure*}[t]
    \includegraphics[width=\textwidth]{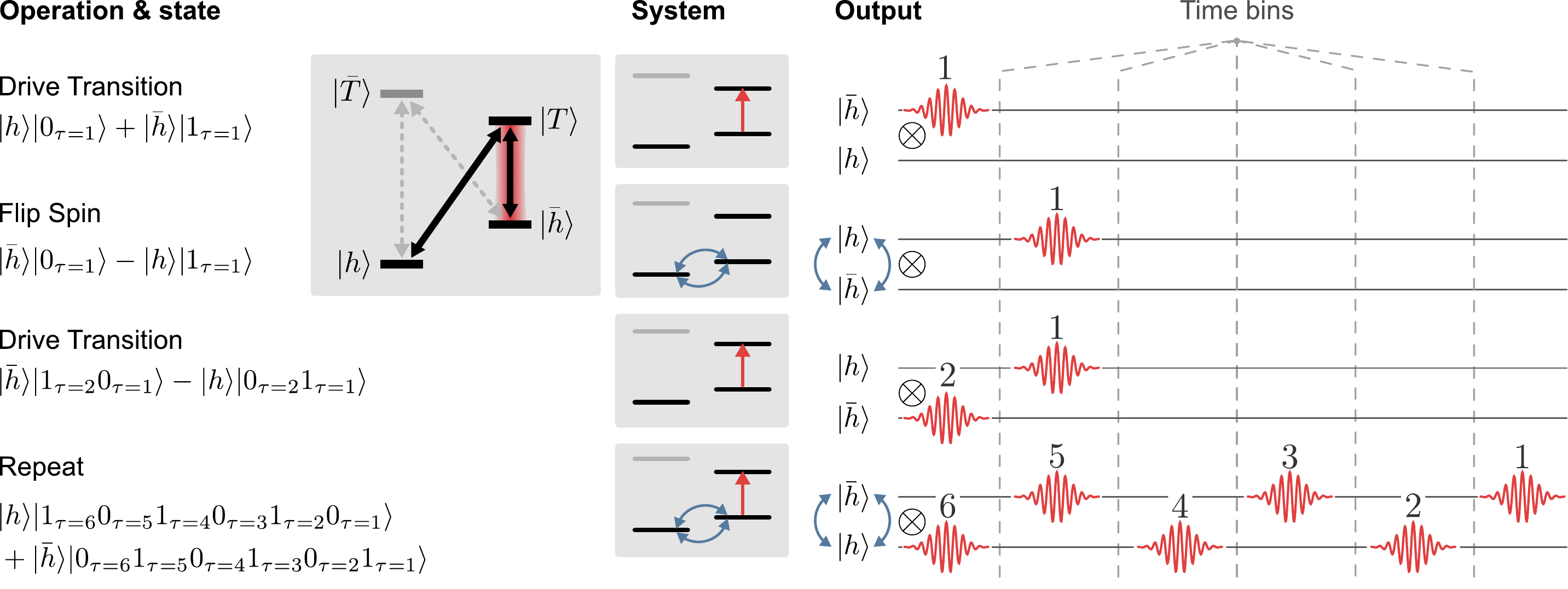}
    \caption{ 	 
    A diagram illustrating our scheme for generating time-bin-encoded multi-photon entangled states. The scheme uses a charged quantum dot in a Voigt geometry magnetic field with a selectively cavity-enhanced vertical transition. The Zeeman-split hole (trion) states are denoted by $\ket{h}$ and $\ket{\bar{h}}$ ($\ket{T}$ and $\ket{\bar{T}}$).
    }
    \label{scheme}
    \end{figure*}

    Time-bin-encoded states are well suited for transmission through optical fibre and integrated waveguide technologies as they are largely immune from decoherence \cite{brendel1999pulsed,marcikic2004distribution}. 

    We propose a modification of the scheme in \cite{lindner2009proposal} to generate photons entangled in the time-bin basis rather than the polarisation basis. Our scheme involves repeatedly resonantly driving a single transition so that all photons are produced with the same frequency and polarisation.

    We propose using a QD with a trapped charge placed in a large Voigt geometry magnetic field. This results in a double lambda system with four transitions of distinct and individually addressable frequencies (Fig. \ref{scheme}). The selective cavity enhancement of a single one of these transitions has been used to demonstrate fast spin preparation, cavity enhanced Raman scattering and the generation of time-bin-encoded single photon states \cite{sweeney2014cavity, lee2018multi}.
    In this scheme, we propose using the cavity enhancement of a single vertical transition to allow a cycling transition suitable for spin state readout and entangled photon generation (Fig. \ref{scheme}).
     
    In addition to increasing the expected number of repeated excitations of the vertical transition resulting in a higher probability of a successful spin readout, cavity enhancement has been shown to improve the coherence properties of the emitted light under resonant excitation. This enables the creation of photons with high indistinguishability that are suitable for HOM interference based operations \cite{bennett2016cavity,giesz2015cavity}.

    Micropillar cavities also allow for increased collection efficiencies relative to planar cavities or non-cavity enhanced systems. Collection efficiencies of up to 79\% have been observed using QDs in micropillar cavities \cite{gazzano2013bright}.

    With the Voigt geometry field allowing spin preparation and manipulation and the cavity enhancement providing us with a cycling transition, our proposal for generating time-bin encoded GHZ states, illustrated in Fig. \ref{scheme}, is as follows:

    \begin{enumerate}
        \item Prepare the trapped hole spin in the $(\ket{h} + \ket{\bar{h}})/\sqrt{2}$ state by performing spin initialisation as in \cite{atature2006quantum} and then using an off-resonant pulse to rotate the spin to the desired state as in \cite{greilich2009ultrafast}.
        \item Resonantly drive the cavity enhanced transition with a $\pi$-pulse to generate a photon in the first time bin conditional on the spin being in the $\ket{\bar{h}}$ state. We are then left with the state $(\ket{h}\ket{0_{\tau=1}} + \ket{\bar{h}}\ket{1_{\tau=1}})/\sqrt{2}$.
        \item Use an off-resonant pulse to flip the spin state of the trapped charge, giving the state $(\ket{\bar{h}}\ket{0_{\tau=1}} - \ket{h}\ket{1_{\tau=1}})/\sqrt{2}$.
        \item Resonantly drive the cavity enhanced transition with a $\pi$-pulse to generate a photon in the second time bin conditional on the spin being in the $\ket{\bar{h}}$ state. This leaves up with the state $(\ket{\bar{h}}\ket{1_{\tau=2}0_{\tau=1}} - \ket{h}\ket{0_{\tau=2}1_{\tau=1}})/\sqrt{2}$.
        \item Another spin flip leaves us with the state $(\ket{h}\ket{1_{\tau=2}0_{\tau=1}} + \ket{\bar{h}}\ket{0_{\tau=2}1_{\tau=1}})/\sqrt{2}$.
        \item Repeating steps 2-5 twice more leaves us with the state $(\ket{h}\ket{1_{\tau=6}0_{\tau=5}1_{\tau=4}0_{\tau=3}1_{\tau=2}0_{\tau=1}} + \ket{\bar{h}}\ket{0_{\tau=6}1_{\tau=5}0_{\tau=4}1_{\tau=3}0_{\tau=2}1_{\tau=1}})/\sqrt{2}$
        \item Then perform a $\pi/2$ spin rotation and resonantly drive the cavity enhanced transition in order to perform a spin readout. This effectively allows us to perform a measurement in the $\ket{\pm}=\frac{1}{\sqrt{2}}(\ket{h}+\ket{\bar{h}})$ basis. Assuming we measure the spin to be in the $\ket{+}$ state, we are left with the photonic state $(\ket{1_{\tau=6}0_{\tau=5}1_{\tau=4}0_{\tau=3}1_{\tau=2}0_{\tau=1}} + \ket{0_{\tau=6}1_{\tau=5}0_{\tau=4}1_{\tau=3}0_{\tau=2}1_{\tau=1}})/\sqrt{2}$. Rewriting this state using a photon in an odd numbered time bin to be a logical 1 and a photon in an even numbered time bin as a logical 0 we have the state $(\ket{111} + \ket{000})/\sqrt{2}$ - a 3 photon GHZ state.
    \end{enumerate}

    For this work we focus on how we could generate a GHZ state as it is conceptually and experimentally simpler than a linear cluster state. In \cite{lindner2009proposal} the authors use a $\pi /2$ y-rotation of the spin between each generated photon in order to create a linear cluster state instead of GHZ state. By making use of the Voigt geometry, we could control the spin state to add the same modification to our scheme. 

    We also note that it might be unexpected for there to be a well defined phase relationship between subsequently generated photons due to the incoherent decay from the excited state. However, for resonant Rayleigh scattered light, the scattered light is coherent with the resonant laser. We expect the resonant Rayleigh scattered component of the light to dominate here due to the cavity enhancement of the transition \cite{bennett2016cavity}, consequently, we expect there to be a well defined phase relationship between the sequentially produced photons.

    In the work below we show that a relatively small Purcell factor of $\sim 5$ is sufficient for a proof-of-principle demonstration, but to generate the desired state deterministically the diagonal transition must be further suppressed relative to the vertical transition. We suggest practical approaches to doing this in the Discussion section.

\section{Experimental work}

    \begin{figure*}[t]
        \centering
        \includegraphics[width=0.9\textwidth]{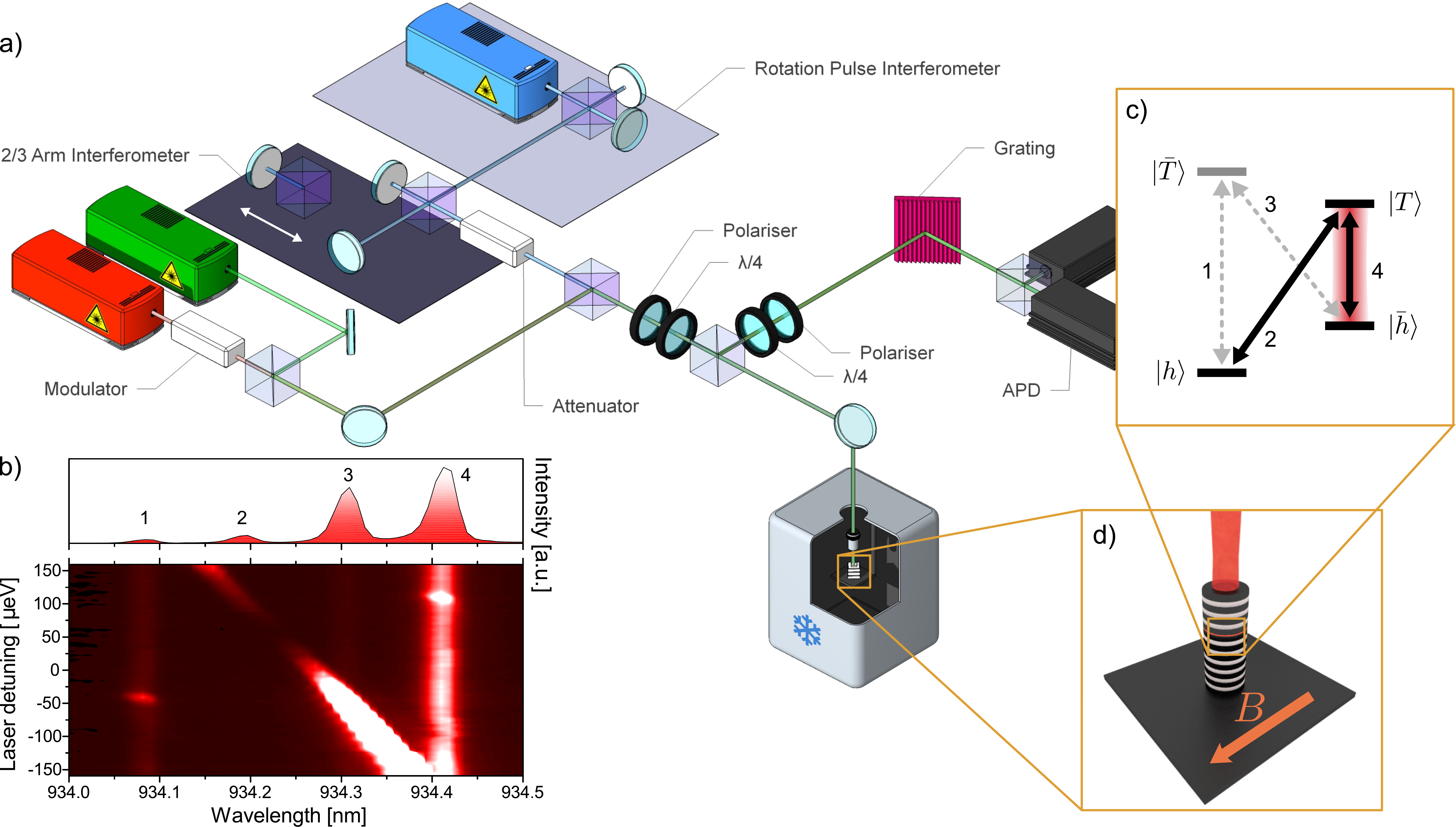}
        \caption{{\bf a)} An illustration of the experimental setup used. The output of the resonant CW laser (red) is modulated using an electro optic modulator. The output is combined on a beam splitter with the output of a pulsed non-resonant laser (green). The output of a pulsed, red-detuned, modelocked Ti:sapphire laser (blue) is directed into an interferometer in order to create the two pulses used perform spin flips. The output of this interferometer is directed into a second interferometer that can have either two or three arms, depending on the pulse sequence required. The output of this second interferometer is combined on a beam splitter with the output of the resonant and non-resonant lasers and focused on the cooled QD-micropillar system via a dark field microscope. The output light is polarisation filtered by the dark field microscope to remove the resonant laser light and a grating is used to spectrally filter the light from the other two lasers. The filtered output light is directed into a pair of Avalanche Photo-diodes (APDs), which enables us to record the arrival time of each recorded photon. {\bf b)} Top: The spectrum of the positive trion transition at 5K under non-resonant excitation in a 9T Voigt geometry magnetic field. Bottom: The spectrum of the positive trion transition under both non-resonant excitation and excitation by a narrow linewidth laser as a function of the narrow linewidth laser detuning from a arbitrary energy between the two diagonal transitions. {\bf c)} The energy level diagram deduced from the results of the experiment shown in (b). {\bf d)} An illustration of the GaAs/AlGaAs micropillar cavity in a Voigt geometry magnetic field.
        }
        \label{setup}
    \end{figure*}

    \subsection{Properties of the quantum dot-microcavity system}

        We used the positive trion transition in this work - the spectrum at 5K under non-resonant excitation in a 9T Voigt geometry magnetic field is shown in Fig. \ref{setup}b. The spectrum indicates that the cavity enhanced transition is Purcell enhanced by a factor of $\sim 5$ and fitting a Lorentzian peak to the cavity mode allows us to determine that the cavity has a Q-factor of $\sim$7500. In order to determine which spectral peak corresponds to each transition, we excited the system using non-resonant light at 850nm and then scanned a narrow linewidth, resonant laser across the central two transitions and observed the corresponding intensity changes in the other transitions (Fig. \ref{setup}b). This allowed us to determine the energy level diagram (Fig. \ref{setup}c).

        As well as being used to investigate the spectrum of the quantum dot, we used the non-resonant laser to probabilistically introduce a hole to the quantum dot - the sample is un-doped and so the quantum dot has a low probability of containing a trapped hole if carriers are not introduced optically. 

        We probed the lifetime of the hole remaining in the dot by probabilistically introducing a hole with a non-resonant pulse, waiting for a given amount of time and then observing the output when driving the cavity-enhanced transition. The results indicate that the hole remains trapped in the quantum dot for times much greater than 50 ns. Future implementations could make use of deterministic charging schemes \cite{lagoudakis2013deterministically, ediger2005controlled, ellis2010electrically,pinotsi2011resonant}.

        A final point to note is that although two transitions are visibly enhanced Fig. \ref{setup}b, only one of the transitions is of interest - the second enhanced transition is due to decay from the $\ket{\bar{T}}$ state, which is not populated during the photon generation scheme.

    \subsection{Manipulation of the trapped hole spin}
\begin{figure*}[t]
            \includegraphics[width=\textwidth]{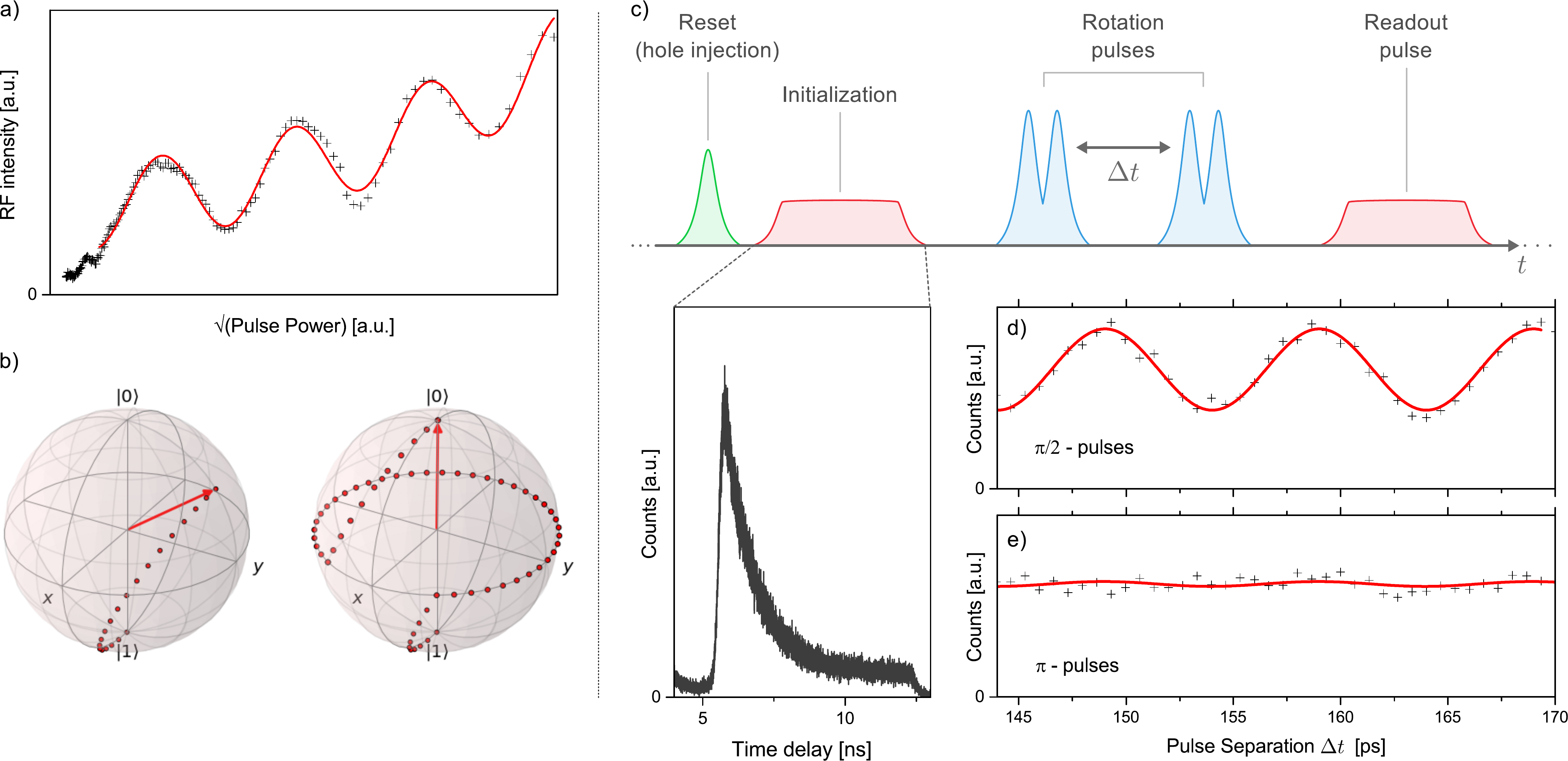}
            \caption{{\bf a)} The measured resonance fluorescence intensity from the readout pulse as a function of the square root of the rotation pulse power. The red line serves as a guide to the eye and is the sum of a sinusoidal function and an exponential. {\bf b)} An illustration of an off-axis rotation of the Bloch vector which misses the pole of the Bloch sphere and does not result in a high-fidelity spin flip. {\bf c)} Left: A two pulse sequence involving two off-axis rotations of the Bloch vector resulting in a high-fidelity spin flip. Right: An illustration of the Ramsey interference measurement pulse sequence and a time-resolved plot of the output during the initialisation pulse showing spin preparation. {\bf d)} The result of the Ramsey interference measurement using $\pi /2$ pulses. {\bf e)} The result of the Ramsey interference measurement using $\pi$ pulses (the two-pulse spin flip scheme).}
            \label{rotations}
        \end{figure*}

        Time resolved measurements of the output from driving the enhanced transition after probabilistically injecting a hole spin shows that the resonance fluorescence intensity decreases exponentially with time - this can be seen in the initialisation pulse of Fig. \ref{rotations}c. This indicates that we can perform spin state preparation, enabling us to prepare the $\ket{h}$ state with high fidelity \cite{atature2006quantum}. We note that cavity enhanced spin preparation in a similar experimental setup has been demonstrated by driving a non-enhanced transition in order to increase the speed of spin preparation \cite{lee2018multi}. However, our choice to drive the cavity enhanced transition allows us to only use a single resonant laser, reducing the experimental complexity at the cost of longer spin preparation times.

        In order to coherently rotate the spin, we use a modelocked Ti:Sapphire laser to produce pulses that are $\sim 6$ ps in length and red-detuned from the transitions. This allows for spin manipulation via the AC Stark effect \cite{greilich2009ultrafast}. In order to demonstrate this, we prepare the system in the $\ket{h}$ state, apply a rotation pulse, and then apply another resonant pulse to serve as a readout pulse - we will only see emission if the $\ket{\bar{h}}$ state has a finite probability of being occupied. Fig. \ref{rotations}a shows the result of varying the rotation pulse power - Rabi oscillations can be seen in the intensity of the readout pulse. As well as observing the expected oscillations, we note that the average intensity measured during the readout pulse tends upwards as the rotation pulse power is increased. We attribute this to nonlinear effects in the fibre altering the spectral and temporal profile of the pulse \cite{kaldewey2017coherent}, which reduces the fidelity of the spin rotation.

        Finally, in order to demonstrate complete control of the spin state, we perform Ramsey interference with the hole spin. An interesting detail here is that, unlike in prior work, a single rotation pulse is not sufficient to flip the spin state. Due to the high magnetic field, the hole spin precession time is $\sim 10$ ps - comparable to the length of the Fourier transform limited rotation pulse, $\sim 6$ ps (prior to entering the fibre). As a result, a single pulse does not perform a rotation about the x (or an equivalent) axis, but about an axis with some z-component - high fidelity spin flips are therefore not possible (Fig. \ref{rotations}b left). To counter this problem, we both lower the magnetic field to 6T to increase the spin precession time and use a two-pulse sequence as in \cite{mizrahi2014quantum}, to flip the spin. This two-pulse sequence allows us to perform a complete $\pi$ rotation as illustrated in Fig. \ref{rotations}b right). As high magnetic fields are required to separate the transitions enough to allow for selective cavity enhancement, the precession time for the trapped spin will be short. Consequently this two pulse rotation scheme is likely to be useful for all realisations of our scheme.

        To observe Ramsey interference using this two-pulse rotation scheme, we  use the pulse sequence illustrated in Fig. \ref{rotations}c. We first use a non-resonant pulse to inject a hole into the QD and then resonantly drive the $\ket{\bar{h}}\rightarrow\ket{T}$ transition to prepare the spin (Reset and Initialization pulses). We then apply the Ramsey pulse sequence, which consists of the two-pulse rotation sequence, followed by a delay, $\Delta t$, followed by a second two-pulse rotation sequence (Rotation pulses). Finally, drive the enhanced transition again to perform a projective measurement - the measurement of a photon is a projective measurement of the $\ket{\bar{h}}$ state.
        Fig. \ref{rotations}d) shows the intensity of the RF from the readout pulse as a function of $\Delta t$, the separation of the rotation pulses.
        When we set the rotation pulse power such that each two-pulse sequence results in a $\pi/2$ rotation of the spin we observe Ramsey fringes with a visibility of 51\%. When we set the rotation pulse power such that each two-pulse sequence results in a $\pi$ rotation, we do not observe Ramsey fringes (Fig. \ref{rotations}e), indicating that we are reliably flipping the spin state as expected. 

        As can be seen in Fig. \ref{rotations}d and \ref{rotations}e, we do not see high Ramsey interference visibilities relative to prior work \cite{press2008complete}. We attribute this to both the slow drift in laser power and intensity over the duration of a measurement and the aforementioned broadening and chirping of the pulse due to non-linear effects in the fibre.

        Finally, we use the Ramsey interference measurement at 6T to extract the $T_{2}^{*}$ time of the hole spin. From an exponential decay fit to the envelope we find that $T_{2}^{*} = 2.445\pm 0.357$ ns, which is in agreement with prior work \cite{sun2016measurement,de2011ultrafast}. Although this is long enough for this experimental demonstration, there are techniques available to extend the coherence time via dynamical decoupling and nuclear field manipulation \cite{press2010ultrafast, malein2016screening}.

\section{Proof-of-principle demonstration}

    \begin{figure}[t]
        \centering
        \includegraphics[width=7cm]{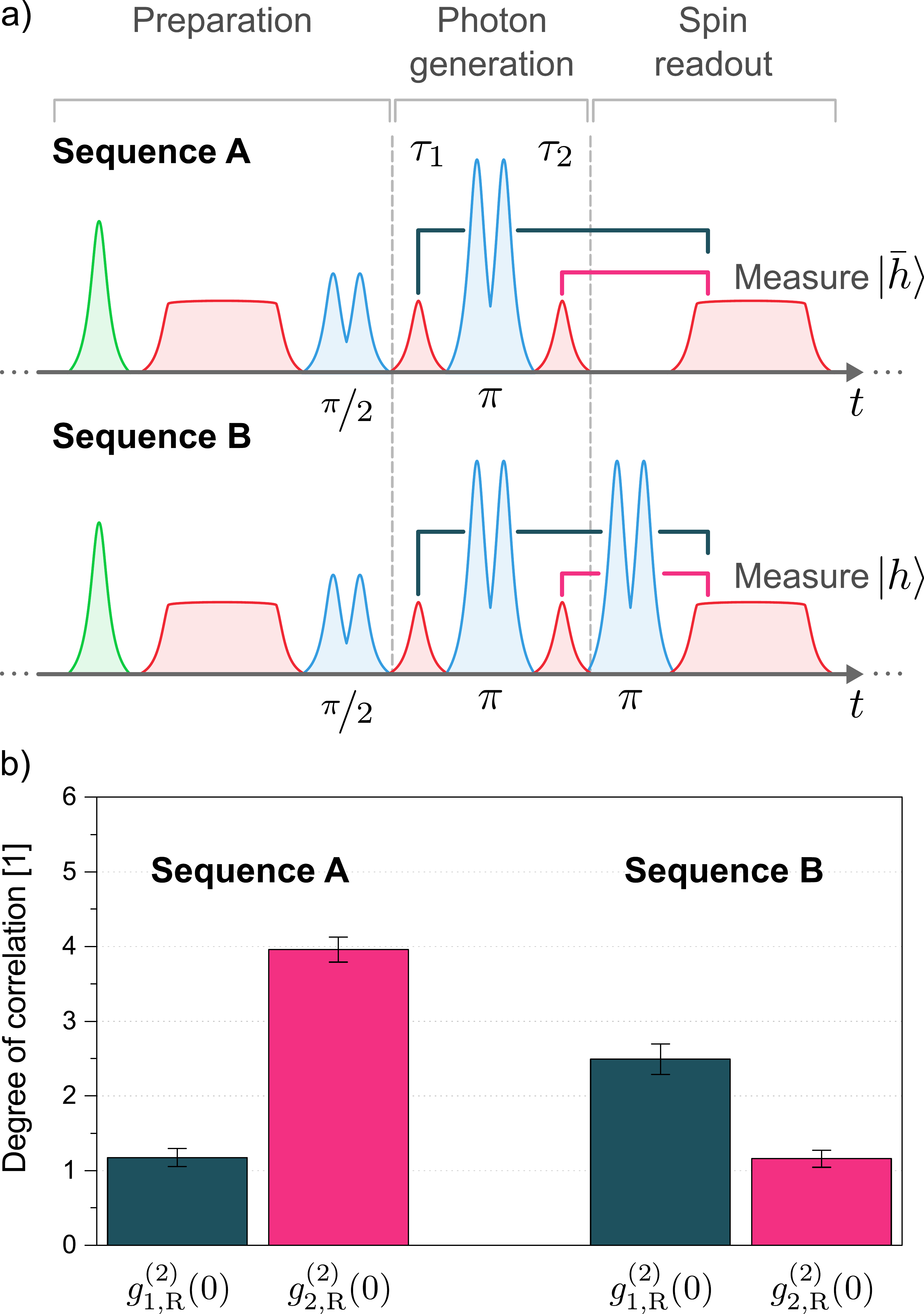}
        \caption{{\bf a)} The pulse sequence to generate a time-bin-encoded photon that is entangled with the state of the hole spin. In sequence A, the readout stage is simply a pulse resonant with the enhanced transition - the measurement of a photon projects the hole spin into the $\ket{\bar{h}}$ state. In sequence B, the combination of a spin flip and the resonant pulse results in the measurement of a photon projecting the hole spin into the $\ket{h}$ state. The coloured lines between the photon generation pulses and the readout pulse indicate the time ranges of the pulse sequence that we use to calulate the degrees of correlation plotted in (b). {\bf b)} Measurements of the degree of correlation between the photon generation pulses and the readout pulse for each pulse sequence.}
        \label{correlations}
    \end{figure}

    In this section, we show that the time bin of the photon is dependent on the measured state of the spin.

    The pulse sequence for generating a time bin encoded photon is shown in Fig. \ref{correlations}a).
    The non-resonant pulse probabilistically injects a hole spin into the quantum dot, then the resonant spin-preparation pulse prepares the spin in the $\ket{h}$ state. A two-pulse sequence is used to perform a $\pi/2$ spin rotation which prepares the spin in a superposition state. Then a sequence of a photon generation pulse, followed by a two-pulse $\pi$ rotation, and a second photon generation pulse is used to generate a photon in the early or late time bin, dependent on the state of the spin. The result of this process should be to generate a spin - time-bin entangled state. Finally, we projectively measure the spin state. To measure the $\ket{\bar{h}}$ state, we use a long resonant pulse to drive the $\ket{\bar{h}}\rightarrow \ket{T}$ transition - measuring a photon here a projects the hole into the $\ket{\bar{h}}$ state. To measure the $\ket{h}$ state, we apply a two-pulse $\pi$ rotation sequence to flip the populations of the $\ket{h}$ and $\ket{\bar{h}}$ states and then measure the $\ket{\bar{h}}$ state as before. Time-tagging each measured photon allows us to investigate the correlations between different events. 

    We use the notation $g^{(2)}_{1,R}(0)$ and $g^{(2)}_{2,R}(0)$ to refer to the degree of correlation between the first photon generation pulse and the readout pulse, and between the second photon generation pulse and the readout pulse respectively. 

    We define the degree of correlation as the measured number of photons recorded from a photon generation pulse conditional on measuring a photon in the readout pulse within the same repetition of the pulse sequence, divided by the average number of these coincidences recorded when we consider photon generation pulses and readout pulses in different repetitions of the pulse sequence.  

    Fig. \ref{correlations}b shows how the degree of correlation for each photon generation pulse changes depending on the spin state that we measure. We see that for sequence A, where we measure the system in the $\ket{\bar{h}}$ state, that if we record a photon produced by the photon generation pulses, there is a probability of $g^{(2)}_{2,R}(0)\,[g^{(2)}_{1,R}(0)+g^{(2)}_{2,R}(0)]^{-1} \approx 0.77$ that the photon will be in the second time bin.
    However, for sequence B, where an additional $\pi$ rotation of the spin after the photon generation step means that we measure the spin in the $\ket{h}$ state, the probability that a photon produced by the photon generation pulses will be measured in the first time bin is  $g^{(2)}_{1,R}(0)\,[g^{(2)}_{1,R}(0)+g^{(2)}_{2,R}(0)]^{-1} \approx 0.68$. We therefore conclude that the photon time bin is dependent on the measured state of the trapped hole spin.

\section{Discussion}

    Our results indicate that, within the experimental imperfections of our setup, the scheme works as intended. There are some modifications required to verify the generation of entangled states and to improve the efficiencies to the degree that this technique can have practical applications. We comment on these here and note that many of the required improvements have been demonstrated experimentally in recent years, making this a promising approach to the generation of photonic states suitable for use in a measurement-based quantum computer.

    In this implementation we used the electro-optic modulation of a CW resonant laser to generate short excitation pulses. These pulses were long relative to the Purcell enhanced decay time meaning that multiple excitations are possible within the length of the pulse, this will result in the output state differing from the intended output state. Future implementations should make use of shorter pulses in order to avoid this problem, however, we believe that the theoretical investigation of the classes of states that could be produced by varying the number of photons generated in each excitation pulse (whether accidental or intentional) would be an interesting avenue for further work.

    In this work, the decay rate of cavity enhanced transition is increased by a factor of $\sim 5$, meaning that the system is $\sim 5$ times more likely to decay vertically, as required, than decaying diagonally. This ratio could be improved by both increasing the Purcell factor or by switching to the Faraday geometry and removing the diagonal transitions. Although the Voigt geometry is typically used when performing spin rotations, spin rotations in the Faraday geometry are possible in principle due to imperfections in the selection rules \cite{de2013ultrafast} and some degree of coherent control of a spin in the Faraday geometry has already been demonstrated \cite{lee2016ramsey}.

    Lastly, the spin used here had a relatively short coherence time - enough for the demonstration but it could become a limiting factor for long pulse sequences. Our scheme involves a $\pi$ pulse half way through, which should increase the effective coherence time via the spin echo effect. As well as spin echo techniques, future work could make use of techniques to reduce the nuclear spin noise such as \cite{ethier2017improving} in order to increase the spin coherence time.

\section{Conclusion}
    We have presented a new scheme for generating entangled, time-bin encoded multi-photon states. Recent work on resonance fluorescence of quantum dots in micropillar cavities have shown photons generated in this manner are highly indistinguishable, so we expect the generated light to be suitable for HOM based fusion operations.
    Experimentally, we have demonstrated complete coherent control of a trapped hole spin via multi-pulse sequences and shown that the time bin of the generated photons is dependent on the measured state of the spin.
    Finally, we suggest improvements that have already been demonstrated with current technology that would improve upon our results in order to generate chains of entangled and indistinguishable photons.

\section*{Acknowledgements}
    The authors acknowledge funding from the EPSRC for MBE system used in the growth of the micropillar cavity. J. L. gratefully acknowledges financial support from the EPSRC CDT in Photonic Systems Development and Toshiba Research Europe Ltd. B. V. gratefully acknowledges funding from the European Union's Horizon 2020 research and innovation programme under the Marie Sk\l{}odowska-Curie grant agreement No. 642688 (SAWtrain).

\section*{Data Access}
    The experimental data used to produce the figures in this paper is publicly available at [Information will be made available on Cambridge Data Repository before publication].

\bibliographystyle{apsrev4-1}
\renewcommand{\bibname}{References}
\bibliography{references}

\end{document}